\documentstyle[prl,aps]{revtex} 
\begin{document} 
\draft 
\preprint{IMSc-97/06/21; hep-th 9707124} 
\title{Black hole fermionic radiance and D brane decay} 
\author{Saurya Das, Arundhati Dasgupta, Parthasarathi Majumdar
and Tapobrata Sarkar\footnote{email: saurya, dasgupta, partha, sarkar@imsc.ernet.in}} 
\address{The Institute of Mathematical Sciences, CIT Campus, 
Chennai (Madras)- 600113, India.} 
\maketitle 
\begin{abstract}
The semiclassical grey-body factor for massless fermion emission from the four dimensional 
black hole described by an ensemble of intersecting triplets of D- 
five-branes is shown to be 
consistent with the (statistical) decay rate of the branes (in the `long' D-string 
approximation) into massless fermionic closed 
string states, subject to assumptions regarding the energy distribution of 
colliding open string states. 
\end{abstract}

Entropies of almost ideal gases of specific BPS and `near'-BPS D brane configurations are 
known to match the Bekenstein-Hawking entropies \cite{mal1} of corresponding black hole 
solutions  around their extremal limit. The instability of (a gas of) excited (non-BPS) D 
branes owing to decay into closed string modes also quantitatively simulates, \cite{cal} 
in the `long'   
string \cite{dm} {\it and} `dilute gas' \cite
{mal2} approximation, Hawking evaporation and absorption rate
\cite{dha} of the 
corresponding black hole. The observed 
similarities, however, do not yet appear to illumine crucial issues of generic black hole 
physics like the causal structure associated with horizon formation, or the information 
loss problem \cite{haw}. Also, disagreements of the rates were
observed in few cases \cite{dis}. Therefore, their scope warrants careful assessment, more so 
because attention thus far has been exclusively confined (with a very recent exception 
\cite{gub}) to emission of bosons \cite{cal,dm,haw,dis,boson,kleb,mal2}. 
The present letter focuses on Hawking radiation of 
a specific four dimensional black hole into 
massless neutral {\it fermions}, in relation to the (statistical) decay of corresponding 
D-brane configurations into massless 
fermionic closed string states. 

We consider the `long' D-string which happens to be the line of intersection of three 
intersecting five-branes in $M$-theory \cite{kleb} toroidally 
compactified to 
four spacetime dimensions. A four dimensional black hole solution to the low 
energy effective theory is given by the metric
\begin{eqnarray} 
ds^2~&=&~-f^{-1/2}hdt^2~+~f^{1/2}\left(h^{-1}dr^2~+~r^2d \Omega^2
\right)~, ~~~\nonumber\\ 
f~&=&~\prod_{i=1}^4 \left( 1~+~{r_i \over r} \right),~~~ h~=~1 - \frac{r_0}{r}~~,
\label{metr}
\end{eqnarray}
where, $r_i$ are related to the four $U(1)$ charges of the black hole, in turn 
proportional respectively to the numbers of the three five-branes under consideration (for 
$i=1,2,3$), and the energy-momentum along the intersection line D-string. The quantity 
$r_0$ is the location of the horizon. An almost ideal (or dilute) gas of these D-brane 
configurations corresponds to the domain $r_0~,~r_4~\ll~r_i~,~i=1,2,3$ \cite{mal2}. We shall 
henceforth confine to this domain. The semiclassical greybody factor for massless 
(chiral) fermion emission from this black hole is calculated from the solutions to the 
Weyl equation in the background (\ref{metr}), in the `near' and the `far' regions with 
respect to the black hole horizon, and subsequently matching these solutions at 
some intermediate region to determine the incoming and absorbed fluxes. The ratio of these 
fluxes yields the required absorption cross section (grey-body factor)
\cite{mal2,unruh}. 

A convenient choice for the local tetrad components appropriate to (\ref{metr}) is given by
\begin{eqnarray} 
e^0_t~&=&~f^{-1/4}h^{1/2}~~,~~ e^3_r~=~f^{1/4}h^{-1/2}, ~~ \nonumber \\
e^1_{\theta}~&=&~f^{1/4}r~~,~~ e^2_{\phi}~=~f^{1/4}~r \sin \theta~,~ \label{tetr}
\end{eqnarray} 
yielding the Weyl equations (for the radial components of the Weyl spinor field, assumed 
left-handed) 
\begin{eqnarray} 
i~r\sqrt{f/h}~\omega~R_1 + r \sqrt{h}~\frac{d R_1}{dr} ~&=&~-\lambda~R_2~, \nonumber \\ 
i~r\sqrt{f/h}~\omega~R_2 - r \sqrt{h}~\frac{d R_2}{dr} ~&=&~\lambda~R_1~, \label{weyl} 
\end{eqnarray}
where $\lambda$ is a constant, and $\omega$ is the frequency associated with the time 
dependence of the solutions, assumed $\exp (-i \omega t)$. 

Near the horizon ($r \rightarrow r_0$), we introduce the variable $z \equiv 1 - r_0/r$ and
approximate $f$ as  
$$f~=~K~\left [(1-z)^4~\sinh^2\sigma + (1-z)^3 \right]~,$$ 
where $K \equiv r_1r_2r_3/r_0^3$, and $r_4 = r_0 \sinh^2 \sigma$ defines $\sigma$. From eq. 
(\ref{weyl}), a second order differential equation is easily deduced for either of the 
radial functions. For $R_1$, we take the ansatz 
$$R_1~=~A~z^m~\left( 1 - z \right)^n~F(z)~,$$
and restricting ourselves to the region
$r\sim r_0$,
we obtain a hypergeometric equation for $F(z)$
\begin{eqnarray} 
z\left(1-z\right)~\frac{d^2F}{dz^2}~&+&~\left[ \left( 2m +
\frac{1}{2}\right) - z\left( 1+ 2m  + 2 n + \frac{1}{2}\right) \right] ~\frac{dF}{dz}~
\nonumber \\
 &-& \left[ 
 \left(m + n + \frac{1}{4}\right)^2~+~ \mu(\omega,
\sigma)\right]F~=~0~~. \label{hyp}
\end{eqnarray}

Here, we have already made the choices,  
$~ m~=~-i\left(\frac{a+b}{2}\right)~,~ n^2~=~\lambda^2~$ with
$a(resp.~ b) \equiv \omega \sqrt{r_1 r_2 r_3 / r_0} ~e^{\sigma}~ (resp.~ 
e^{-\sigma})~$.
In the regime $a \approx b$ corresponding to $\sigma\sim 0$, the  
function $\mu(\omega, \sigma)=-1/16 +\imath (a + b)/8$
and hypergeometric function relevant to (\ref{hyp}) is given by $F(\alpha, \beta, \gamma; 
z)$ where, $\alpha\approx-i\frac 34 (a+b) +n + \frac{1}{2}~,~\beta\approx-i\frac14 (a+b)+n 
~,~\gamma=-i(a + b) + \frac12 ~.$ The physical significance of this regime may be open to 
question since it corresponds to $r_4 \ll r_0$, and since all semiclassical considerations 
pertain to behaviour outside the blackhole horizon. However, we include this for 
completeness. 

In the other regime where $\sigma\geq 1$ corresponding to $a>b$, we
have $\mu(\omega, \sigma)~=~-\left[1/4 - \imath (a-b)/2 \right]^2$.
The solution for $R_1(z)$ in the near zone can thus be 
shown to be 
\begin{equation} 
R_1^{near} = A z^{ -i(a+b)/2}~(1-z)^n~F \left(\alpha,\beta,\gamma;z \right)~,~
\alpha=-ia +n + \frac12,~ \beta=-ib +n ,~ \gamma=-i\left(a +
b\right) + \frac{1}{2}~.  \label{hypsol1} 
\end{equation}
For large $\sigma$, $b\approx 0$, and  
parameters are $\alpha = -ia + n + \frac12,~\beta = 
n~,~\gamma = -ia+ \frac12~. $ 
This can be derived alternatively by using the Newman-Penrose
formalism. In either case, in the limit $z \rightarrow 
0~,~R_1^{hor}(z) = A z^{-i (a+b)/2}~.$

The near zone solutions obtained above are to be matched by extrapolation to the regime 
$z \rightarrow 1$ to the small distance limit of the solution in the far zone ($f 
\rightarrow 1~,~h \rightarrow 1$). The latter solution can be shown to be given
in terms of the Whittaker function \cite{grad} as
\begin{equation}
R_1^{far}(r)~~=~~\frac{B}{\sqrt{\omega r}}~W_{\frac12,n}(\omega r)~.\label{far}
\end{equation}
Using the small distance limit of the Whittaker function \cite{grad}, and matching with 
the near solutions yields the ratio of the constants $A$ and $B$, in absolute value, to be
\begin{equation}
\left |{B \over A}\right |~=~\frac{1}{\sqrt{2}}(2\omega r_0)^n \left|{\Gamma(n) \over 
\Gamma(2n)} {{\Gamma(\gamma)~ \Gamma(\gamma-\alpha -\beta)} \over  
{\Gamma(\gamma-\alpha) ~\Gamma(\gamma-\beta)}} \right|~~. \label{bbya}
\end{equation}

Now, with the conserved fermionic flux given by  
\begin{equation} 
{\cal F}~~=~~|R_1|^2 - |R_2|^2~,  
\end{equation} 
the absorption cross section of interest\footnote{The contribution of $R_2$ is negligible.} 
can be expressed as 
\begin{equation}
\sigma_{abs}~=~{{2 \pi} \over {\omega^2}}~{{\cal F}_0 \over {\cal F_{\infty}}}~=~{\pi 
\over \omega^2}~ \left | {A \over B} \right|^2~~,\label{grf}
\end{equation}
where, ${\cal F}_0~(resp. {\cal F}_{\infty})$ is the flux entering the black hole horizon 
(resp. flux arriving from past infinity). Thus, the Hawking radiation rate for fermions 
is given by 
\begin{equation}
\Gamma_H~=~ ~{\pi \over \omega^2}~ \left | {A \over B} \right|^2~(e^{2\pi(a+b)} 
+1)^{-1}~{d^3k \over (2\pi)^3}~. \label{hawk}
\end{equation} 

For $a \approx b$, using eq. (\ref{bbya}) and appropriate values of the parameters 
$\alpha,~ \beta~and ~\gamma$ with $n=-1$ (which corresponds
to s-wave solution for the Weyl fermion) we get
\begin{equation}
\Gamma_H~=~\frac14 \pi^2 r_0^2 (a+b) [\frac14 + \frac{9}{16} (a+b)^2]~\left(e^{\frac{\pi}{2} 
(a+b)} -1 \right)^{-1} \left( e^{\frac32 \pi(a+b)} + 1\right)^{-1}~
\frac{d^3k}{(2\pi)^3} \label{smal}
\end{equation}
For $a> b$, we get, for $n=-1$,
\begin{equation}
\Gamma_H~=~4\pi^2 r_0^2 {b(\frac14 + a^2) \over {(e^{2\pi a}+1)~(e^{2\pi b} -1)}} {d^3k 
\over (2\pi)^3}~. \label{big}
\end{equation}
Also, with $n=-1$ one obtains for $a \gg b$
\begin{equation}
\Gamma_H~=~2\pi r_0^2~\left[ \frac14~+~a^2 \right]~\left ( e^{2\pi
a} ~+~1 
\right)^{-1}~ \frac{d^3k}{(2\pi)^3},~\label{gp}
\end{equation}
which essentially is the solution found in ref. \cite{gub}.
This result can be derived using the alternative
Newman-Penrose formalism. Also, taking $\omega \rightarrow 0$,
we reproduce the general result in \cite{dmg}.

Massless closed string fermionic modes (gravitinos) are emitted from the decay of excited 
intersection D-strings (corresponding to the black hole considered above) via collision of 
bosonic and fermionic massless open string modes propagating on them. 
Gravitinos with vector polarization in the
compact direction emerge as chiral spin-1/2 particles 
akin to 
those considered above. The calculation of the D-string decay rate into such particles to 
lowest order in string coupling entails evaluating the disc amplitude with a fermionic and a 
bosonic vertex operator inserted at the boundary and a closed string fermionic vertex 
operator in the bulk. Since 
the latter can be decomposed into open string bosonic and fermionic vertex operators, the 
requisite matrix element may be written, with suitable identifications \cite{gar}, as (in 
the light cone Green Schwarz formalism \cite{gg})
\begin{equation}
{\cal A}(p_1,p_2,p_3,p_4)~=~V_{CKG}^{-1}\int \prod_{i=1}^4 dz_i~< 
{V_F(1)~V_B(2)~[V_B(3)~V_F(4)~+~V_B(4)~V_F(3)]}>~, \label{matr}
\end{equation}
where,
\begin{eqnarray}
V_B(n) & \equiv & V_B(\zeta_n,p_n,z_n) \equiv \zeta^I_n B^I_n(p_n,z_n)~e^{ip_n \cdot 
X(z_n)}~\nonumber \\
V_F(n) & \equiv & V_F(u_n, p_n, z_n) \equiv \left( u^a F^a(p_n,z_n) + u^{\dot a} F^{\dot 
a}(p_n,z_n) \right) e^{p_n.X(z_n)}~, \label{vert}
\end{eqnarray}
with $n=1, \dots,4~I,a,{\dot a}=1, \dots, 8$. In this 
formulation, matrices $M^{IJ}$ and ${\cal M}^{ab}~,~{\cal M}^{{\dot a}{\dot b}}$ 
have been introduced a la' \cite{gg} to account for the boundary conditions 
associated with the D-string under consideration. 

Rather than explicitly computing the rhs in (\ref{matr}), one can take recourse to the 
Ward identities corresponding to the surviving spacetime supersymmetries for the D-string 
\cite{gg}, which may be expressed schematically as
\begin{equation}
V_{CKG}^{-1}\int \prod_{i=1}^4 dz_i~< \left [\eta Q~,~V_F(1) V_B(2) V_B(3) V_B(4) 
\right ] >~=~0~.\label{ward}
\end{equation}
This enables us to express the desired amplitude purely in terms of the bosonic 
amplitude calculated for instance in ref. \cite{hash}. 
Here we merely estimate the energy dependence of the 
amplitude to check consistency with the semiclassical results. The amplitude for 
two open string bosons on the D-string fusing into a graviton with polarization $\epsilon$ 
transverse to the D-string is given by \cite{hash}
\begin{equation}
{\cal A}_B~\sim~{\Gamma(-2t) \over \left(\Gamma (1-t)\right)^2} t^2 \left ( 
\zeta_1\cdot \epsilon \cdot \zeta_2 \right)~.\label{bos}
\end{equation}
Using (\ref{ward}) one obtains, once again for vector polarization components of the 
gravitino being transverse to the D-string, the amplitude (suitably covariantized)
\begin{equation}
{\cal A}_F~\sim~{\Gamma(-2t) \over \left(\Gamma (1-t)\right)^2} (2t) \left ({\bar u}_1 
\zeta_2 \cdot \gamma \xi_3 \cdot k_2 + \ldots \right )~,\label{Fermi}
\end{equation}
where, $\xi_3$ is the polarization vector-spinor of the emitted fermion. (Note we have written only a representative polarization
term to illustrate momentum dependence).
Strictly speaking, the physical region of interest is for small values of 
$t~\sim~\omega^2$; it is easy to see that in this domain, $|{\cal A}_F|~\sim~\omega$.  

In contrast to standard computations, as is commonly accepted \cite{cal}, a decaying gas 
of D-strings does not afford the usual facility of {\it preparing} an initial state. 
The standard procedure of {\it averaging} over initial states for `unpolarized' initial 
states must therefore be replaced by a {\it summation} over all possible initial momentum 
distributions. If we assume that the gas in question is an ensemble in thermal 
equilibrium, it follows that the colliding open string Bose and Fermi modes will also 
`thermalize' according to their natural statistics, with distributions $\rho_B(\omega,T_B)$ 
and $\rho_F(\omega,T_F)$ respectively, where, the temperatures $T_B, T_F$ may not be 
equal. The total decay rate for the D-string with appropriate
phase space factors is then given by
\begin{equation} 
\Gamma_D~\approx~\omega~\rho_F(T_F)~\rho_B(T_B)~\frac{d^3k}{(2\pi)^3}~ . 
\end{equation}  
Comparison with our semiclassical analysis, in the limit of
$\omega\rightarrow 0$, (upto coefficients)
leads to, from equations (\ref{smal}) and 
(\ref{big}) 
\begin{eqnarray}
T_B~&=&~{ \omega \over {\pi (a+b)}}~~,~~T_F~=~{\omega \over {3 \pi(a+b)}}~~, a \approx  
b~\nonumber\\ 
T_B~ &=&~{ \omega \over {4\pi b}}~~,~~T_F~=~{\omega \over {4\pi a}}~~,~~a> b~. \label{temp}
\end{eqnarray}
We have not yet completed the task of computing the temperatures $T_B~,~T_F$ from the 
statistical mechanics of massless open string states on the D-string, but hope to report 
it in the near future. 
Observe that in the case $a > b$, the results in 
(\ref{big}) can also be understood in terms of left-moving and right-moving distributions 
of open-string modes on the D-string. This, as mentioned, is not manifest in (\ref{gp}). However why the  effective fermi temperature is lower valued, is not clear.
Also, while our choice of $n=-1$ appears to be an admissible value for $s$-wave 
scattering, other solutions also ought to be explored for novel possibilities. The calculation of the exact string amplitude is
being pursued to see whether the decay rates match exactly.
\noindent

{\bf Acknowledgements :}
We would like to thank T. Jayaraman and B. Sathiapalan
for useful discussions. S.D. and T.S. would like to thank 
Mohan Narayan for fruitful discussions.

\end{document}